\documentclass[11pt]{article}
\usepackage{amssymb}
\usepackage{pst-all}
\usepackage{pstricks}
\usepackage{pstcol,pst-fill,pst-grad}
\usepackage{amsfonts}
\usepackage{graphicx}
\usepackage{amsmath}
\usepackage[normalem]{ulem}
\usepackage{multicol}
\usepackage{tikz}
\usepackage{enumerate}

\usetikzlibrary{matrix}
\RequirePackage{mathrsfs} \RequirePackage[sc]{mathpazo}
\RequirePackage{wasysym} \RequirePackage{setspace}
\newcommand{\be}{\begin{equation}}
\newcommand{\ee}{\end{equation}}
\newcommand{\bea}{\begin{eqnarray}}
\newcommand{\eea}{\end{eqnarray}}
\input{tcilatex}
\begin{document}

\date{}
\title{Quantum Information and Elementary Particles}
\author{ Adil Belhaj$^{a}$ \thanks{%
belhaj@unizar.es}, Salah Eddine Ennadifi$^{b}$ \thanks{%
ennadifis@gmail.com}, \\
\\
{\small $^{a}$ LIRST, Polydisciplinary Faculty, Sultan Moulay Slimane
University }\\
{\small B\'eni Mellal, Morocco } \\
{\small $^{b}$ LHEP-MS, Faculty of Science, Mohammed V University, Rabat,
Morocco} }
\maketitle

\begin{abstract}
Motivated by  string theory and standard model physics, we discuss
the possibility of other particles-based quantum information.  A
special  attention is put on the consideration of the graviton in
light of the gravitational wave detection. This may offer a new take
in approaching quantum information using messenger particles. The
construction is readily extended to higher dimensional qubits where
we speculate on possible connections with open and closed string
sectors in terms of quiver and graph theories, respectively. In
particular, we reveal that the  vectorial qubits could be associated
with skeleton diagrams considered as  extended quivers.

\end{abstract}

\newpage

\section{Introduction}

Quantum Information Theory (QIT) has attracted recently much
attention mainly in connection with many subjects including
condensed matter, particle physics, string theory, graph theory,
black holes and communication tools \cite{1,2,3,30,4}. This theory
is considered as a bridge between computer science and quantum
mechanics. In particular, it is based on a fundamental component
known by qubit. This piece has been investigated using certain
mathematical operations corresponding to tensor-product of Hilbert
vector spaces. Precisely, qubits have been extensively dealt with by
applying various methods including type II superstrings, D-branes
and graphic representations. Concretely, a nice link between the
stringy black holes and qubit systems have been studied by
exploiting the compactification scenario. In these activities, it
has been shown that the supersymmetric STU black hole obtained from
the type II superstrings has been related to 3-qubits using the
hyperdeterminant concept\cite{1}. This correspondence has been
enriched by many generalizations associated with toric and
superqubit calculations. All these works have lead to a
classification of qubit systems in terms of black objects in type II
superstrings using  D-brane physics.

 Recently, a strong
interest has been devoted to  apply graphs in different aspects of
 QIT. In particular, models have been developed from Adinkras
explored in the study of the supersymmetric representation theory.
These graphs have been used to classify a class of qubits in terms
of extremal black branes resulted from abelian gauge theories
involving  photons which are a particular case of the fundamental
 messengers, belonging to the electromagnetic
interaction. This comes up with the idea that other interaction
messengers could be considered in quantum information processing.

The main objective of this paper is to discuss the possibility of
other messenger particles-based QI by combining symmetry and
geometry in the context of string theory and standard model of
particle physics (SM). Motivated by the recent detection of
gravitational waves (GWs), a particular attention is devoted to  the
consideration of the graviton in QI. This may offer a new take on
studying QI using the known elementary particles. The construction
is readily extended to higher dimensional qubits where we speculate
on possible connections with open and closed string sectors in terms
of quiver and graph theories, respectively. Concretely,   we show
that the vectorial qubits  could be associated with  skeleton
diagrams considered as extended quivers in string theory and related
topics.

The present work is organized as follows: In section 2, we outline
how the bosons, according to SM, can be implemented in QI.  Section
3 concerns the spin two particle providing gravitational qubits
supported by the recent detection of GWs. In section 4,  multiple
qubits are speculated  in terms of extended quivers and graphs. The
last section contains some concluding remarks and perspectives.

\section{Vectorial  quantum information }

In this section,  we bridge a relation between QIT and bosonic
fields appearing either in SM or in string theory. To do  so, let us
recall that the qubit is a two-state system \cite{5}. A
superposition of a single qubit is generally given by the following
Dirac notation

\begin{equation}
|\psi \rangle =a_{0}|0\rangle +a_{1}|1\rangle \text{ \ \ }  \label{eq1}
\end{equation}%
\ where $a_{i}$ are scalars belonging to a field $F$ satisfying the
normalization condition

\begin{equation}
|a_{0}|^{2}+|a_{1}|^{2}=1.  \label{eq2}
\end{equation}%
The analysis can be extended to more than one-qubit, given in (\ref{eq1}),
which have been used to discuss entangled states. For $n$-qubits, the
general state has the following form

\begin{equation}
|\psi \rangle =\Sigma _{i_{1}...i_{n}=0,1}a_{i_{1}...i_{n}}\left\vert
i_{1}...i_{n}\right\rangle  \label{eq3}
\end{equation}%
where $a_{i_{1}...i_{n}}$ satisfy the normalization condition

\begin{equation}
\Sigma _{i_{1}...i_{n}=0,1}|a_{i_{1}...i_{n}}|^{2}=1.  \label{eq4}
\end{equation}%
It has been remarked that, in almost all quantum activities, the
single qubit has been associated with the polarization of the photon
producing photonic qubits. The later has been considered as the
channel for sending QI in four dimensional universe. A close
inspection, however, in particle physics and string theory, shows
that the photon, $\gamma $, is not the only particle having
two-polarization states. In particular, there are other bosons which
could play a similar role as photons. Indeed, the QI could be
transferred using other bosonic particles living in four dimensions.
These particles belong to three types.

\begin{enumerate}
\item The first type contains the vector particles appearing in  the three
fundamental interactions: electromagnetic, weak and strong interactions.
These particles are noted by $\gamma ,\;W^{+,-,0},$ and$\;g^{a}$ ($%
a=1,\ldots ,8$) respectively. It is recalled that $g^{a}$ are the eight
gluons associated with the Cartan decomposition of $su(3)$ Lie algebra \cite%
{6}. This class of particles can be supported from the fact that in four
dimensions the massless vector bosons have two-polarization states. It is
recalled that on-shell massless vectors in $d$ dimensions have $d-2$ degrees
of freedom. In fact, it has been shown that one degree of freedom is removed
by the field equation and another one by the gauge condition.

\item The second type is the Higgs boson $H$ involving two degree of freedom
\cite{7}. It is recalled that this particle is the only known discovered
scalar associated with many sectors in SM and beyond.

\item The third type is the massless graviton $g$ which in four dimensions
involves also two-polarization states. This can be motivated from
the fact that in such a dimension a GW has two-polarization states.
This can be seen also
form the fact that a massless graviton in $d$ dimensions have $\frac{%
(d-2)(d-1)}{2}-1$ degrees of freedom. It is mentioned also that this
particle appears naturally in the quantization of the closed string
theory. Recent detection of GWs might be exploited to support the
idea that graviton could play a relevant r\^ole in communication and
quantum  information systems.\newline
\end{enumerate}

In this vision, it seems important to support this statement inspired by
particle physics and string theory \cite{8}. Investigations show that we can
provide some arguments by combining symmetry, geometry, and physics. At
first sight, it has been not clear how to approach such a problem. However,
it has been remarked that symmetry in physics can be exploited to discuss
the present proposition. Focusing on the state algebraic equation and
keeping the two dimensional vector space analysis associated with two-state
polarization in particle physics, one has only a freedom to inspect the
field $F$ on which the vectors are built. An inspection shows that this
filed involves certain symmetries similar to the ones appearing in particle
physics and related topics including SM and string theory.

In what follows, we will show that these symmetries are hidden in the
division structure of the field $F$ \cite{9}. In this way, the state $|\psi
\rangle $ physics should be invariant by the Lie symmetry corresponding to $F
$. In photonic qubit, the state is invariant by the phase transformation $%
e^{i\alpha }|\psi \rangle $ where $\{e^{i\alpha }\}$ form the U(1)
symmetry. The latter is the electromagnetic interaction symmetry
corresponding to the two-polarization state  photons in QIT. This
phase transformation, which produces projective spaces required by
the probability condition, is associated with the Bloch sphere in
photonic qubits. According to fundamental interactions, one may
propose the similar vectors of the weak interaction corresponding to
SU(2) transformation $X|\psi \rangle $, where $X $ is a SU(2)
element. It has been inspected that the field $F$ should be
associated with the quaternionic projective geometries required by
the probability condition. The involved bosons are massive leading
to more than two-polarization states, which could be evinced from
the present discussion. However, the gluons $g^{a}$ associated with
the su(3) Lie symmetry can be implemented in QI.  It is observed the
octonionic nature of SU(3) color. Indeed, it is recalled that
$G_{2}$ is the automorphism group of the octonions \cite{10}. The
latter acts transitively in the $S^{6}$ sphere of unit imaginary
octonions implying that the $S^{6}$ acquires a
quasi-complex structure assured by the sequence of inclusion $\mbox{SU(3)}%
\rightarrow G_{2}\rightarrow S^{6}$. This can be considered as an
octonionic representation of SU(3) symmetry. The appearance of it in
QI is the more remarkable one. We remember the original group SU(3)
flavour of Gell-Mann, mixing the first three quarks flavors (up u,
down d and strange s). It is recalled that Gell-Mann and Fritzsch
introduced also SU(3) as the gauge (color) group of the strong
interactions, giving rise to QCD\cite{11}. Howover, the connection
with QI needs more reflexing ideas based on QCD activities.

\section{Gravitational qubits}

Having discussed the vector bosons in QI, one may consider other
particles supported by models going beyond  SM. It is worth noting
that the bosonic  vectorial spectrum can be derived from open string
sector and related topics. While, the graviton comes from the closed
string sector and should considered differently. As mentioned in the
previous section, one may consider the graviton in light of the
confirmed GWs observation \cite{12}.
In general relativity (GR) \footnote{%
For simplicity, natural units with $\hbar =c=1$ are considered.}, the metric
$g_{%
\mu
\upsilon }$ describing the geometry of spacetime relates the spacetime
coordinate $dx^{\mu }$ to the spacetime interval $d\ell ^{2}$ through

\begin{equation}
d\ell ^{2}=g_{%
\mu
\upsilon }dx^{\mu }dx^{\nu }.  \label{eq5}
\end{equation}%
By the fact of the extreme weakness of the GWs at the earth, the metric can
be approached  as that of the Minkowski flat metric $\eta _{%
\mu
\upsilon }$ by ignoring the background curvature.  In the case of
small perturbations,  the equation of GR can be linearized. Thus,
these perturbations can be approximated as the sum of the flat
metric and a small perturbation induced by the GWs as

\begin{equation}
d\ell ^{2}=\eta _{%
\mu
\upsilon }+h_{%
\mu
\upsilon }  \label{eq6}
\end{equation}%
where $\left\vert h_{%
\mu
\upsilon }\right\vert \ll 1$ is the GWs induced small perturbation. The
small changes in the spacetime interval $\delta \ell ^{2}$ can be expressed
by the small perturbations and the spacetime coordinate like

\begin{equation}
\delta \ell ^{2}=h_{%
\mu
\upsilon }dx^{\mu }dx^{\nu }  \label{eq7}
\end{equation}%
where $\pm \delta \ell ^{2}$ represent the small contractions and dilations
changes in the length $\ell $ of the spacetime structure. In the most useful
choice, the solution of the GR equation becomes a system of linear
equations, in particular a system of wave equations corresponding to a three
dimensional wave equation traveling at the speed of light $c=1$ \cite{13}.
With the symmetry of $h_{\mu \upsilon }$ and in the sinusoidal case, the
physical part of these waves can be written here in the $z$-direction by the
wave equation

\begin{equation}
h_{\mu \upsilon }=\delta \ell \epsilon _{%
\mu
\upsilon }\cos \left( \omega t-kz\right)   \label{eq8}
\end{equation}%
where $\omega $ and $k$ being the angular frequency and the wave vector
respectively. The elements $\epsilon _{%
\mu
\upsilon }$ are the so-called unit polarization tensors $\epsilon _{%
\mu
\upsilon }^{+}$, $\epsilon _{%
\mu
\upsilon }^{\times }$ with the signs $+$,$\times $ saying that there
exist just two possible independent polarization states such as
$\delta \ell \epsilon _{\mu \upsilon }=\delta \ell ^{+,\times
}$which represent the strain amplitudes of each polarization.
Naively, the absence of continuous gauge symmetry associated with
the gravity and completeness of division algebra push us to think
about other symmetries associated with the metric calculations.
Concretely, the corresponding qubit should be associated with a real
geometry involving discrete gauge symmetries. It is noted that such
symmetries have been investigated in the context of gravitational
theories supported by black hole activities. A straightforward way
to look for such a symmetry is to  handle  the algebraic qubit
equation. This suggests that the continuous symmetries of SM can be
replaced by the $Z_{2}$ gauge symmetry in the gravitational qubits.
Going through this argument, it is possible to convince ourself to
propose such a such symmetry for developing real Bloch spheres as
geometric representations of gravitational qubits. This goes beyond
the projective spaces appearing in photonic qubits. In this way,
two-polarization states can be associated with a real vector space
derived from the real condition $a_{0}^{2}+a_{1}^{2}=1$, providing
the circle
equation. This equation is invariant under the $Z_{2}$ symmetry acting as $%
a_{i}\rightarrow -a_{i}$. The reader might be puzzled with this
argument. To make it more transparent, it would be important to
provide a  contact with concrete observations. In this way, it is
intersecting to note that GWs have now been confirmed by the recent
successful detection in the LIGO experiment and hence the detection
of the graviton becomes now more likely and thus graviton-based QI
seems possible.

\section{Higher dimensional qubits}

Before ending this statement, let us discuss higher dimensional
qubits in the present language. It has been remarked that this is
not an easy task. Borrowing ideas from the relation between D-branes
in type II superstrings and gauge theories, one may approach higher
dimensional vectorial  qubits using the quiver method associated
with open string sector \cite{14}. In string/M-theory, this method
has been used to study  four dimensional D-brane gauge theories
derived from the compactification either on singular Calabi-Yau
manifolds or on $G_{2}$ manifolds. The
 matter content
of the resulting models can be obtained from the geometric
deformations and the topological invariant data of the internal
manifolds. These models are usually refereed to quiver gauge
theories. In this way, the physical content of a model with several
continuous gauge group factors can be encoded in a quiver. As in
graph theory, the quiver is formed by nodes and edges. For each
node, one associates a gauge factor $G_{\ell}$, where $\ell$
indicates the node on which the physical information is put.
However, edges between two nodes are associated with the charged
matter transforming either in
bi-fundamental or fundamental representations of the gauge group $%
G_{quiver}=\prod_{nodes \;\ell}G_{\ell}$. In such theories, various
models of such quivers have been elaborated and built in connection
with toric graphs and Dynkin diagrams. Roughly speaking, the
$n$-qubits, discussed in the present work, may be associated with a
quiver theory with $G_{quiver}=G^{n}$ where $G $ is the continuous
symmetry corresponding to  polarization states of the vector
particles obtained from the open string sector. In this way, the
multiple photonic qubits corresponds to a quiver diagram of a
particular symmetry defined by $n$ pieces of the U(1) gauge group
associated with the electromagnetic interaction. In string theory,
the quantum states are $2^{n}$ ways of $n$ D-branes that wrap $n$
distinguishable cycles in the internal Calabi-Yau space. In this
graph langauge, the nodes correspond to qubits and edges exhibit the
presence of entanglement between    a pair of qubits $\{i_1,i_2\}$.
Precisely, the nodes which are not connected by edges correspond to
no entangled
 two qubits. It is recalled that the bipartite entanglement of $i_1$ and $i_2$
qubits  is given in terms of the 2-tangle $\tau_{i_1i_2}$
\begin{equation}
\tau_{i_1i_2}= 4|deta_{i_1i_2}|^2.
\end{equation}
In quiver theory, the edges however  are associated with
$Q_{i_1i_2}$ bi-fundamental matter charged  under the $G\times G$
gauge symmetry. It seems that the connection between $
\tau_{i_1i_2}$  and  $Q_{i_1i_2}$  is possible using some form
applications.  The link  pushes  us   to  discussion higher
dimensional qubit systems.  The curiosity shows that these systems
should be dealt with  differently since they will be associated with
matter having more than two charges going beyond the bi-fundamental
fields. To incorporate such a  matter, the quivers should be
replaced by skeleton diagrams, as done in  \cite{140}. In these
diagrams, the gauge factors are represented by edges and matter  by
nodes. This can be done by thinking on a  dual operation  which
exchanges  the r\^ole of  the node and the  edges in the  quiver
method.  It has been observed that  the  skeleton  diagrams  involve
polyvalent vertices being connected to more than two other ones. In
string theory,  the associated graphs have been used  to represent
indefinite Lie algebras generalizing the finite and affine
symmetries explored in the geometric engineering method of four
dimensional gauge theories obtained from type II superstrings
compactified on Calabi-Yau manifolds\cite{141}.  It is remarked that
the tri-vertex appears in the finite so(8) Lie algebra while the
tetra-vertex arises naturally in the affine so(8) Dynkin diagram.
Roughly, the poly-vertex of order $n$ ($n$-vertex), in skeleton
activities, should correspond  to  a field $Q_{i_1\ldots i_n}$
transforming in the poly-fundamental representation of $G\times
G\times \dots\times G$ going beyond bi-fundamental matters
associated with 2-qubits. Based on these observations, one might
attempt to link $n$-tangle $\tau_{i_1\ldots i_n}$ in $n$-qubit
systems  with the poly-vertex  of order $n$  in skeleton diagrams
\begin{equation}
\tau_{i_1\ldots i_n}\leftrightarrow |det Q_{i_1\ldots i_n}|^2.
\end{equation}
It is believed  that the  link  is at least not a direct one, and
several aspects of it need more reflexing ideas.

Developed investigations in generalized quivers could provide lights
on building interesting qubits by  considering different symmetries
placed at edges. This involves the symmetries appearing in particle
physics including the one of SM involving photons and other similar
particles. We refer to these systems as mixed qubits which would  be
dealt with using quiver and skeleton  methods.
\newline Investigations show that multiple gravitational qubits
could be associated with four dimensional multi graviton theories
being non-interacting spin two massless fields described by a sum of
Pauli-Fierz actions. The latter is a linearized form of a sum of
Einstein-Hilbert actions which is the free action for GR. This model
has been approached using graph theory \cite{15}. The latter  can be
exploited to study QI in which the gravitational  $n$-qubits are
associated with graph containing $n$ nodes and a sent of edges
connecting such nodes. The graph provides a mathematical framework
for approaching the quantum states and related concepts including
the entanglement between gravitational  qubits. Using similarities
between multi-gluons and multi-gravitons scattering, we  expect that
graph theory can be used to study multi gravitational   qubits in
terms of the physics of spin two massless fields in four dimensions.

\section{Conclusion and perspectives}

In this work, we have investigated the possibility of implementing
messenger particles in QIT by combining symmetry and geometry in the
context of open and closed string sectors. In the open string sector
associated with vector particles, we have approached qubits in terms
of the corresponding division algebra. In particular, we have
proposed how one can go beyond photonic qubits by exploring the role
of fundamental interaction of particle physics in communication
systems. The closed string sector has been introduced to discuss the
 gravitational qubits supported by the recent observation of GWs.
Multi-qubit systems have also speculated using quiver and graph
methods in terms of D-brane and multi-graviton theories in four
dimensions. In particular, we have shown that the vectorial qubits
could be linked with skeleton diagrams explored in the study of four
dimensional gauge theories.

This work comes up with many open questions. The natural question is
to think about communications in terms of GWs. In this way, the
geometric modifications of the physical space-time could be
relevant. It is worth noting that quantum geometry is the
appropriate modification of ordinary classical geometry to make it
suitable for describing the physics of quantum gravity associated
with string theory. It would therefore be of interest to try to use
quantum information geometry in terms of size and sheep
modifications. We believe that this approach deserves to be studied
further.  Moreover,  the  direct link between vectorial qubits and
poly-matter in  skeleton diagrams would  be investigated  elsewhere
in future.
\\
\\
 { \bf Acknowledgements:} The authors are grateful to their
families for support. \newline

\end{document}